\documentclass[twocolumn,showpacs,preprintnumbers,amsmath,amssymb]{revtex4}
\usepackage{graphicx}
\begin{document}

\title{Electron-like and photon-like excitations in
an ultracold Bose-Fermi atom mixture}

\author{Yue Yu$^{1,2}$ and S. T. Chui$^2$}
\affiliation{1. Institute of Theoretical Physics, Chinese Academy
of Sciences, P.O. Box 2735, Beijing 100080, China}
 \affiliation{2. Bartol
Research Institute, University of Delaware, Newark, DE 19716, USA}
\date{\today}
\begin{abstract}
We show that the electron-like and photon-like excitations may
exist in a three-dimensional Bose-Fermi Hubbard model describing
ultracold Bose-Fermi atom mixtures in optical lattices. In a Mott
insulating phase of the Bose atoms, these excitations are
stabilized by an induced repulsive interaction between 'electrons'
if the Fermi atoms are nearly half filling.  We suggest to create
'external electric field' so that the electron-like excitation can
be observed by measuring the linear density-density response of
the 'electron' gas to the 'external field' in a time-of-flight
experiment of the mixture. The Fermi surface of the 'electron' gas
may also be expected to be observed in the time-of-flight.
\end{abstract}

\pacs{03.75.Lm,67.40.-w,39.25.+k,71.30.+h}

\maketitle

Ultracold atoms in optical lattices have offered a highly tunable
platform to study various physical phenomena which may not be
definitely clarified in condensed matter systems \cite{exa}. On
the other hand, new systems which may not be realized in condensed
matter content are presented, for example, mixtures of Bose-Fermi
atoms as constitution particles. Experimentally, the Bose-Fermi
atom mixtures in optical lattices have been realized for
$^{87}$Rb-$^{40}$K \cite{rbk}, and $^{23}$Na-$^6$Li \cite{nali}.

Microscopically, these mixtures may be described by a Bose-Fermi
Hubbard model \cite{aie}. The constitution particles in this model
are spinless boson($a_i$) and fermion ($f_i$) with $i$ the lattice
site index. In this Letter, we consider three-dimensional cubic
lattices. We will show that, in high temperatures, there are only
excitations of these constitution particles. We call this a
confinement phase. The system undergoes a phase transition, in
certain critical temperature, to  a uniform mean field (UMF) state
in which electron-like and photon-like excitations emerge if the
fermion occupation is nearly half filling \cite{wen}. At the exact
half filling, this UMF state turns to a long range ordered state,
the checkerboard crystal of 'electrons' \cite{comp}. We show that
this UMF state can only be stable if the bosons are in a Mott
insulator (MI) ground state.

We suggest an experiment to create an 'external field' by changing
the depth of the fermion's optical potential \cite{change}. The
response function of the mixture to the external field may be
measured by the density distribution image in a time-of-flight of
the mixture cloud. The behavior of the response functions may be
used to identify the electron-like excitation. We expect the
'electron' Fermi surface can be observed by the time-of-flight
experiment, which has been used to observe the Fermi surface of
the pure cold Fermi atoms \cite{kms}.

The Bose-Fermi  Hubbard Hamiltonian we are interested in reads
\begin{eqnarray}
H&=&-\sum_{\langle ij\rangle}(t_Ba^\dagger_ia_j+t_Ff^\dag_i
f_j)-\sum_i (\mu_Bn^a_i+\mu_Fn^f_i)\nonumber\\
&+&\frac{U_{BB}}2\sum_i n^a_i(n^a_i-1)+U_{BF}\sum_i n^a_in^f_i,
\label{sh}
\end{eqnarray}
where the lattice spacing $\lambda/2$ is set to unit. ( We also
set $\hbar=c=1$.) $n^a_i=a^\dag_ia_i$ and $n^f_i=f^\dag_if_i$.
$t_B$ and $t_F$ are the hopping amplitudes of the boson and
fermion between a pair of nearest neighbor sites $\langle
ij\rangle$. $\mu_B$ and $\mu_F$ are chemical potentials. And
$U_{BB}$ and $U_{BF}$ are the on-site interactions between bosons,
and between boson and fermion. In this work, we use
$U_{BB}>U_{BF}>0$ although this is not necessary in general. The
microscopic calculations of these model parameters in terms of the
cold atom mixture have been established, e.g, in Ref. \cite{aie}.

To deduce the low energy theory in strong couplings, we will use
the slave particle technique, which has been applied to the cold
boson system \cite{d,yu,lu}. In the slave particle language, the
Hamiltonian reads $H=H_2+H_4$ where $H_2$ and $H_4$ are the
two-operator and four-operator terms, respectively. Namely,
\begin{eqnarray}
&&H_2=-\sum_i\sum_\alpha[\mu_B\alpha( n^\alpha_{c,i}+
 n^\alpha_{h,i})+\mu_F n^\alpha_{c,i}]\\
&&+\frac{U_{BB}}{2}\sum_i\sum_\alpha\alpha(\alpha-1)
(n_{c,i}^\alpha+n_{h,i}^\alpha)+U_{BF}\sum_i\sum_\alpha \alpha
n_{c,i}^\alpha,\nonumber
\end{eqnarray}
and
\begin{eqnarray}
H_4&=&-\sum_{ \langle
ij\rangle}t_F\sum_{\alpha,\beta}c^\dagger_{\alpha,i}h_{\alpha,i}
h^\dagger_{\beta,j}c_{\beta,j}\nonumber\\&-&\sum_{
 \langle
ij\rangle}t_B\sum_{\alpha,\beta}\sqrt{\alpha+1}\sqrt{\beta+1}\\
&(&h^\dagger_{\alpha+1,i}h_{\alpha,i}+c^\dagger_{\alpha+1,i}c_{\alpha,i})
(h^\dagger_{\beta,j} h_{\beta+1,j}+c^\dagger_{\beta,j}
c_{\beta+1,j}),\nonumber
\end{eqnarray}
where $n^\alpha_{c,i}=c^\dagger_{\alpha,i}c_{\alpha,i}$ and
$n^\alpha_{h,i}=h^\dagger_{\alpha,i}h_{\alpha,i}$. We explain
briefly the derivation of this slave particle Hamiltonian. The
state configurations at an arbitrary given site consists of $\{
|\alpha,s\rangle~|~\alpha=0,1,2,...~;~s=0,1 \}$ where $\alpha$ and
$s$ are the boson and fermion occupations, respectively. The Bose
and Fermi creation operators can be expressed as
$a^\dag=\sum_\alpha\sqrt{\alpha+1}[|\alpha+1,0\rangle\langle
\alpha,0|+|\alpha+1,1\rangle\langle \alpha,1|], f^\dag=\sum_\alpha
|\alpha,1\rangle\langle\alpha,0|$. The mapping to the slave
particle reads $|\alpha,0\rangle\to h^\dag_\alpha,
|\alpha,1\rangle\to c^\dag_\alpha$ . We call $c_\alpha$ the
composite fermion (CF) \cite{comp} and $h_\alpha$ the slave boson.
The normalized condition
$\sum_\alpha(|\alpha,0\rangle\langle\alpha,0|+|\alpha,1\rangle\langle\alpha,1|)=1$
implies a constraint $\sum_\alpha(n^\alpha_{c}+n^\alpha_{h})=1$ at
each site. The slave particles arise a $U(1)$ gauge symmetry $
 c_{\alpha,i}\to
e^{i\varphi_i}c_{\alpha,i},h_{\alpha,i}\to
e^{i\varphi_i}h_{\alpha,i}. $  The global $U(1)$ symmetry $
c_{\alpha,i}\to e^{i\alpha\varphi}c_{\alpha,i}, h_{\alpha,i}\to
e^{i\alpha\varphi}h_{\alpha,i}$ reflects the particle number
conservation. Since the slave particle technique essentially works
in the strong coupling region, we focus on $U_{BB}/(6t_B)\gg 1$.

There are two types of four slave particle terms in $H_4$, the
$t_B$-terms and $t_F$-terms. We first neglect the $t_B$-terms in
the mean field level of the CF. To decouple the $t_F$-terms , we
introduce Hubbard-Stratonovich fields
$\hat\eta^{c,h}_{\alpha\beta,ij}$ and
$\hat\chi^{c,h}_{\alpha\beta,ij}$. The partition function is given
by $Z=\int D\hat\chi D\hat\eta D\bar c Dc D\bar h Dh D\Lambda
~e^{-S_{eff}}$ where the effective action reads
\begin{eqnarray}
&&S_{eff}[\hat\chi,\hat\eta,c,h,\Lambda]=\int_0^{1/T} d\tau
\biggl\{H_2+\sum_ii\Lambda_i\nonumber\\&&+\sum_i\sum_{\alpha}\biggl[\bar
c_{\alpha,i}(\partial_\tau-i\Lambda_i)c_{\alpha,i}+\bar
h_{\alpha,i}(\partial_\tau -i\Lambda_i)h_{\alpha,i}\biggr]
\nonumber\\
&&+\sum_{ \langle
ij\rangle;\alpha,\beta}t_F[\hat\eta^h_{\beta\alpha,ji}
(\hat\chi^c_{\alpha\beta,ij}-\bar
c_{\alpha,i}c_{\beta,j})-\hat\chi^c_{\alpha\beta,ij}\hat\chi^h_{\beta\alpha,ji}\nonumber\\
&&+\hat\eta^c_{\alpha\beta,ij}(\hat\chi^h_{\beta\alpha,ji} -\bar
h_{\beta,j}h_{\alpha,i})]+t_B~{\rm terms}\biggr\}, \label{ea}
\end{eqnarray}
 where $\Lambda_i$ is a Lagrange multiplier for
the constraint $\sum_\alpha(n^\alpha_{c,i}+n^\alpha_{h,i})=1$.
Rewriting
$\hat\eta^{c,h}_{\alpha\beta,ij}=\eta^{c,h}_{\alpha\beta,ij}e^{i{\cal
A}_{ij}}$,
$\hat\chi^{c,h}_{\alpha\beta,ij}=\chi^{c,h}_{\alpha\beta,ij}e^{i{\cal
A}_{ij}}$ and $\Lambda_i=\Lambda+{\cal A}_{0,i}$, ${\cal A}_{0i}$
and ${\cal A}_{ij}$ are $U(1)$ gauge field corresponding to the
gauge symmetry \cite{tj}. Before going to a mean field state, we
first require the mixture is stable against the Bose-Fermi phase
separation. For example, it was known that the mixture is stable
in the MI phase if $4\pi t_F\sin(\pi n^f) U_{BB}>U_{BF}^2$
\cite{aie}. Near the  half filling of Fermi atoms, this condition
is easy to be satisfied. Fixing a gauge, $\hat
\eta_{\alpha\beta,ij}^h\approx \eta^h_{\alpha\beta,ij}$, $\hat
\chi_{\alpha\beta,ij}^h\approx \chi^h_{\alpha\beta,ij}$ and
$\Lambda_i\approx\Lambda$ (which is a saddle point value of
$\Lambda_i$). This is corresponding to a mean field approximation.
The effective mean field action is given by
\begin{eqnarray}
S_{MF}&=&iN_s\beta\Lambda+\int d\tau\biggl\{\sum_{\langle
ij\rangle}(
t_F\sum_{\alpha\beta}\chi^c_{\alpha\beta,ij}\chi^h_{\beta\alpha,ji})\\
&+&\sum_{i\ne j}\sum_{\alpha\beta}(\bar
c_{\alpha,i}(D^{\alpha\beta}_c)^{-1}_{ij}c_{\beta,j}+\bar
h_{\alpha,i}(D^{\alpha\beta}_h)^{-1}_{ij}h_{\beta,j})\biggr\}\nonumber
\end{eqnarray}
where $(D^{\alpha\beta}_c)^{-1}_{ij}=(\partial_\tau
-\mu^\alpha_F)\delta_{ij}\delta_{\alpha\beta}
-t_F\sum_{\vec\tau}\chi^h_{\alpha\beta,ji}\delta_{j,i+\vec \tau}$
with $\mu^\alpha_F=\mu_B\alpha+\mu_F
-\frac{U_{BB}}2\alpha(\alpha-1)-U_{BF}\alpha-i\Lambda$ and
$(D^{\alpha\beta}_h)^{-1}_{ij}==(\partial_\tau
-\mu^\alpha_B)\delta_{ij}\delta_{\alpha\beta}
-t_F\sum_{\vec\tau}\chi^c_{\alpha\beta,ji}\delta_{j,i+\vec \tau}$.
with $\mu_B^\alpha=\mu_B\alpha
-\frac{U_{BB}}2\alpha(\alpha-1)-i\Lambda $. ($\vec \tau$ is the
unit vector of the lattice.)

The mean field equations are $ \chi^c_{\alpha\beta,ij}=\langle
c^\dag_{\alpha,i} c_{\beta,j}\rangle=T\sum_n
D^{\alpha\beta}_{c,ij}(p_n), \chi^h_{\alpha\beta,ij}=\langle
h^\dag_{\alpha,i} h_{\beta,j}\rangle=T\sum_n
D^{\alpha\beta}_{h,ij}(\omega_n)$ where
$D^{\alpha\beta}_{c,ij}(p_n)=\int d\tau
e^{ip_n\tau}D^{\alpha\beta}_{c,ij}(\tau)$ and
$D^{\alpha\beta}_{h,ij}(\omega_n)=\int d\tau
e^{i\omega_n\tau}D^{\alpha\beta}_{h,ij}(\tau)$ with $p_n$ and
$\omega_n$ the Fermi and Bose frequencies. Near the critical point
where $\chi^{h,c}_{\alpha\beta,ij}\approx 0$, one can expand
$D_{i\ne j}^{\alpha\beta}$ by $\chi^{h,c}_{\alpha\beta,ij}$. For
$\alpha\ne\beta$, the mean field equations are
$\chi_{\alpha\beta,ij}^c=t_F\chi_{\beta\alpha,ji}^h
\frac{n_c^\alpha-n_c^\beta}{\mu_F^\alpha-\mu_F^\beta}$ and
$\chi_{\alpha\beta,ij}^h=t_F\chi_{\beta\alpha,ji}^c
\frac{n_h^\alpha-n_h^\beta}{\mu_B^\alpha-\mu_B^\beta}$ with
$n_h^\alpha=[e^{-\beta\mu_B^\alpha}- 1]^{-1},n_c^\alpha=[e^{-\beta
\mu_F^\alpha}+ 1]^{-1}$. The solutions of these mean field
equations can only exist in the weak coupling limit, i.e.,
$\mu_{F,B}^\alpha\ll t_F$: $T_c^{\alpha\beta}=t_F$. Therefore, in
the parameters we are considering, $\chi_{\alpha\beta,ij}^{h,c}=0$
for $\alpha\ne\beta$. For $\alpha=\beta$, the mean field equations
are $ T^\alpha_c\chi_{\alpha\alpha,ij}^c
=T^\alpha_F\chi_{\alpha\alpha,ji}^h$ and
$T^\alpha_c\chi_{\alpha\alpha,ij}^h
=T^\alpha_B\chi_{\alpha\alpha,ji}^c$ with $
T^\alpha_{F,B}=t_Fn^\alpha_{c,h}(1\mp n^\alpha_{c,h})$. The
critical temperatures are then given by
$
T_c^{\alpha}=\sqrt{T^\alpha_FT^\alpha_B}.
$
Below $T_c^\alpha$, the minimized free energy including variables
$\chi_{\alpha\alpha}^{h,c}$ is $ F\propto
-t_F\sum_\alpha(\tau^\alpha )^2\frac{(S^\alpha_2)^2}{S^\alpha_4}$
where $\tau^\alpha=\frac{T^\alpha_c-T}T$ and $S^\alpha_2$,
$S^\alpha_4$ are the fractions of non-zero terms corresponding to
the second order and four order of $|\chi_{\alpha\alpha}|$
\cite{il}. The optimal $|\chi^c_{\alpha\alpha}|^2\propto
\tau^\alpha S^\alpha_2/S^\alpha_4$.

We consider an integer boson filling $n^a=$1 for $\bar U_{BB}\gg
1$. Since $n^{\alpha\ne 1}_h$ and $n^{\alpha\ne 1}_c$ in this
region are very small, there are no solutions of the mean field
equations with $T^{\alpha\ne 1}_c\geq 0$ for the critical
temperature equations. Thus, all slave bosons and CFs with
$\alpha\ne 1$ are confined. Only $T^1_c>0$ can be found. Below
$T_c^1$, the CF $c_1$  and the slave boson $h_1$ are deconfined in
the mean field sense.  In the inset of Fig. 1, we plot $T_c^1$ for
a set of given parameters. The curve $T_p(U_{BB})$ is
corresponding to $\mu^1_F(T_p)=0$, i.e., the effective chemical
potential of $c_1$ vanishes.

We now go to concrete mean field solutions and focus on the near
half filling of the fermions. Because we work in a
three-dimensional lattice, the flux quanta passing a cubic cell
are zero. Thus, there is no a flux mean field state. (The flux
mean field state may exist in a two-dimensional Bose-Fermi Hubbard
model.) Two possible solutions are the dimer and uniform phases.
In a cubic lattice, each lattice site has six nearest neighbor
sites. A pair of slave boson and CF located at the nearest
neighbor sites may form a bond. The dimer phase means for any
given site, only one bond ended at the site is endowed with a
non-zero $\chi_{\alpha\alpha}^{h,c}$ and other five bond carry
$\chi_{\alpha\alpha}^{h,c}=0$. In the uniform phase, each bond is
endowed with the same real values of $\chi^{c,h}_\alpha$ if a CF
(i.e., $a_i^\dag f_i^\dagger|0\rangle=c^\dag_{1i}|{\rm
vac}\rangle$) is surrounded by six slave bosons (e.g.,
$a_j^\dag|0\rangle=h^\dag_{1j}|{\rm vac}\rangle$). In the fermion
half filling, this is a CF checkerboard crystal state \cite{comp}.
Slightly away from the half filling, this is a UMF state with
$h_1$-$c_1$ bond. In the uniform phase, the dispersions of the CFs
and slave bosons are $ \xi_\alpha^{c,h}(k)=t_F|\chi_\alpha^{h,c}|
|\sum_i\cos k_i|$.

 At the
fermion half filling, the mean field free energy favors for the
dimer state. However, as we shall see soon, in the Mott insulator
phase of the bosons, the boson hopping term ($t_B$-term) we have
neglected at the mean field level will contribute a nearest
neighbor repulsion between CFs or slave bosons if $U_{BB}>U_{BF}$.
This repulsion potential will raise the energy of dimer phase
while the uniform phase is not affected. Thus, for our purpose, we
focus on the uniform phase below.

In the mean field approximation, we neglect the boson hopping term
and the gauge fluctuations ${\cal A}_{0i}$, ${\cal A}_{ij}$ which
must be considered if the mean field state is stable. We first
deal with the $t_B$-term. Introduce a Hubbard-Stratonovich field
$\Phi_i=\sum_\alpha\sqrt{\alpha+1}[c^\dag_{\alpha+1,i}c_{\alpha,i}
+h^\dag_{\alpha+1,i}h_{\alpha,i}]$ to decouple $t_B$-term.
$\Phi_i$ may be thought as the order parameter field of the Bose
condensation. The phase diagram of the boson may be determined by
minimizing the Landau free energy associated with the order
parameter $\langle 0|\Phi_i|0\rangle$. The vanishing of the
coefficient of the second order term in the free energy gives the
phase boundary \cite{d,yu}. It has to point out that here the
fluctuation ${\cal A}_{0i}$ has been neglected and a cut-off to
the type of the slave bosons has to be introduced. However, the
experience to work out the pure Bose phase diagram showed that
these approximations could be acceptable \cite{d,yu,lu}.

As expected, the phase diagram of the constitution boson consists
of the Bose superfluid (BSF), the normal liquid and the MI, in
which the MI phase only exists in the zero temperature and an
integer boson filling factor . In the inset of Fig. 1, we show the
boson phase diagram in the same parameters as those in the CF mean
field phase diagram. In the Bose condensate, the boson number in
each site is totally uncertainty. This means the vanishing bond
number or $S_2\approx S_4\approx0$. Thus, the mean field state is
not stable in the BSF. In the MI phase of the bosons, on the other
hand, the boson number in each site is exactly one for $n^a=1$.
Thus, the mean field states may be stable. Furthermore, the
$t_B$-term induces a nearest neighbor interaction between CFs or
slave bosons. This may be seen by taking the $t_B$ term as a
perturbation if $U_{BB}$ and $U_{BB}-U_{BF}$ are much larger than
$t_B$. To the second order of the perturbation, the $t_B$-term
contributes an effective repulsive interaction between the CFs
 or slave bosons in the nearest neighbor sites
\begin{eqnarray}
{\cal J}=\sum_{\langle ij\rangle}Jn^1_{ci}n^1_{cj}+{\rm const}=
\sum_{\langle ij\rangle}Jn^1_{hi}n^1_{hj}+{\rm const'},
\label{int}
\end{eqnarray}
with $J=\frac{16 t_B^2 U_{BF}^2}{U_{BB}(U_{BB}^2-U_{BF}^2)}$. For
the dimer state, this repulsive potential contributes an energy
$J/2$ to a pair of adjacent bonds. For the UMF state, if the
fermion is in half filling, the checkerboard distribution of the
fermions (then CFs) makes ${\cal J}$ no contribution to the
energy. For a doping $\delta$, i. e., slightly away from the half
filling, the energy raises a small amount of the order $J\delta$.
Thus, the UMF state minimizes this induced nearest neighbor
repulsion.

We have shown the mean field states are not stable in the BSF.
Therefore, we focus on the MI phase of the bosons  below and
examine the gauge fluctuations. The zeroth component ${\cal
A}_{i0}$ restores the exact constraint of one type of slave
particles per site while ${\cal A}_{ij}$ restores the gauge
invariance of the effective action (\ref{ea}). These fluctuations
may destabilize the mean field state. To see the stability of the
UMF state against the gauge fluctuation, one should integrate out
$h_1$ and $c_1$. In the long wave length limit ($k\to 0$),
integrating out $h_1$ first and keeping the Gaussian fluctuations
of the gauge field, an effective action in continuum limit is
given by
\begin{eqnarray}
&&S[c_1,A_\mu]=T\sum_n \int d^3k\biggl[\bar c_1
(ip_n-\mu_c^1+ie_0A_0)c_1 \label{action}\\
&&+\frac{1}{2m_c}\bar c_1(k_a+ie_0A_a)^2 c_1 +\frac{1}2 A_\mu
A_\nu\Pi^B_{\mu\nu}(k,\omega_n)\biggr],\nonumber
\end{eqnarray}
where $e_0=\sqrt{\frac{J}{4\pi}}$, $A_\mu={\cal A}_\mu/e_0$,
$m_c\sim 1/(t_F\chi^c_1\delta)$ is the effective mass of CF. The
coupling constant $2\pi e_0^2=J/2$ is a small quantity means that
the Gaussian approximation to the gauge field is reasonable.
$\Pi^B_{\mu\nu}$ is the response function by integrating over
$h_1$. The density-density response function is given by
$[\Pi^B_{00}(\omega,k)]^{-1}=[\Pi^{B(0)}_{00}(\omega,k)]^{-1}+V(k)$
where $V(k)=J\sum_a(1-k^2/2)$ is the Fourier component of the
interaction (\ref{int}) and $\Pi^{B(0)}_{00}(\omega,k)$ is the
free boson response function. The current-current response
function has a form
$\Pi^B_{ij}=(\delta_{ij}-k_ik_j/k^2)\Pi^B_L+(k_ik_j/k^2)\Pi^B_T$.
In the long wave length limit, the transverse part
$\Pi^B_T=-\frac{\omega^2}{k^2}\Pi^B_{00}$. The longitudinal part
$\Pi^B_L(k,\omega) \propto |\langle 0| h_1|0\rangle|^2 $ as
$k,\omega \to 0$. However, out of the BSF phase, $h_1$ does not
condense. Thus, $ A_\mu$ is a transverse field and the action
(\ref{action}) is very similar to electron coupled to a photon
field. We, therefore, call $c_1$  and $A_\mu$ the electron-like
and photon-like excitation, respectively. Since the hopping of the
Bose atom in the optical lattice is short range, the repulsive
interaction between 'electrons' is also short range. If it was
possible to design the hopping of bosons
$t_{B,ij}\propto1/\sqrt{|{\bf i}-{\bf j}|}$, the interaction
between 'electrons' would be purely coulombic, i.e., $V(k)\propto
1/k^2$. The stability of the UMF state against to the gauge
fluctuations may be seen after integrating out $c_1$ field. This
leads to a CF response function
$[\Pi^F_{\mu\nu}(\omega_n,k)]^{-1}=[\Pi^{F(0)}_{\mu\nu}(\omega_n,k)]^{-1}
+[\Pi^{B}_{\mu\nu}(\omega_n,k)]^{-1}$. The UMF is stable when
$\Pi^F_{\mu\nu}(0,0)>0$. In the long wave length limit, it is
positive if $6J>\frac{\pi^2}{m_c k_F^1}$ where $k_F^1$ is Fermi
momentum of the CF. Since $1/m_c\propto t_F \delta$, this
condition holds only when the fermion filling is slightly away
from the half filling.

\begin{figure}
\begin{center}
\includegraphics[width=7cm]{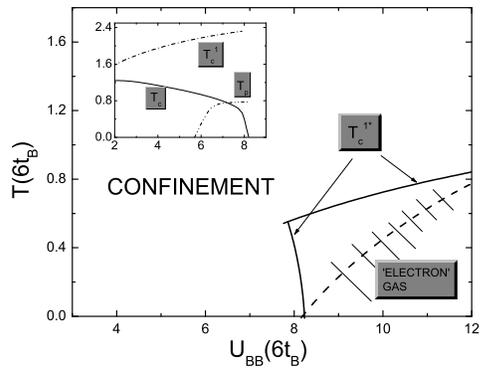}
\end{center}
 \caption{The phase diagram of the CFs in the $U_{BB}$-$T$ plane for
  $n^a=1$, $U_{BF}=U_{BB}/2$,
 $t_F/6t_B=\sqrt{60}$, $n^f=0.55$. The solid curves
 are the critical temperature $T^{1*}_c$ after considering
 the gauge fluctuation and the shrink of the CF Fermi surface.
 }
 \label{1}
\end{figure}

We have shown the stability of the UMF state near the fermion half
filling when the bosons are in the MI phase. We may figure out the
phase diagram of the CF in Fig. 1. The mean field phase transition
temperature $T_c^1$ is suppressed greatly to $T_c^{1*}$ which is
determined by the $T_c$ and $T_p$. The dash line is the estimated
crossover line from the 'electron' gas (the MI of bosons) to the
Fermi liquid of the constitution Fermi atoms (the normal liquid of
bosons as the incompressibility of the MI is gradually
disappearing).

 The experimental implications
of the UMF phase are discussed as follows. We consider the
'electron' response to an external 'electric' field, 'made' by a
change of the lattice potential of the Fermi atoms. This technique
has been used to study the excitation spectrum of atom superfluid
in optical lattices \cite{change}. This disturbs the density of
Fermi atoms with $H'=-\sum_i n^f_i\varphi_i$. In a time-of-flight
experiment, the difference between disturbed and undisturbed
fermion densities by external field is given by $n_f({\bf
r})-n^0_f({\bf r})=(\frac{m_f}t)|\tilde w_f({\bf k}=\frac{m_f{\bf
r}}t)|^2\delta n_f({\bf k}=\frac{m_f{\bf r}}t)$ where $n_f({\bf
r})$ is the image after the time-of-flight and $n_f({\bf k})$ is
the Fourier component of $n_i^f$; $t$ is the flying time, $\tilde
w_f$ is the Fourier component of the fermion Wannier function and
$m_f$ the Fermi atom mass. If $T>T_c^{1*}$, the density response
of the system is simply given by the free fermion one, and then
$\delta n_f({\bf k})\propto\Pi^{F(0)}_{00}$. When the bosons are
in the MI phase ($n^a_i=1$), $n^f_i=n^1_{ci}$ implies $ \delta
n^f({\bf k}) =\delta n^1_{c}({\bf k})$. Thus, for $T<T^{1*}_c$,
especially below the dash line in Fig. 1, $\delta n_f({\bf
k})\propto \Pi_{00}({\bf k},0)$ with $
\Pi_{00}^{-1}=(\Pi^{F(0)}_{00})^{-1}+(\Pi^{B(0)}_{00})^{-1}+V(k)$.
Then, this difference between the response functions may be
measured in experiment.
 A better experimentally measurable quantity is the
visibility ${\cal V}=\frac{n_f({\bf r}_{\rm max})-n_f({\bf r}_{\rm
min})}{n_f({\bf r}_{\rm max})+n_f({\bf r}_{\rm min})}$ where ${\bf
r}_{\rm max}$ and ${\bf r}_{\rm min}$ are chosen such that the
Wannier envelop is cancelled \cite{vi}. The difference between the
disturbed and undisturbed visibility may directly correspond to
the response function because $n_f({\bf r}_{\rm max})+n_f({\bf
r}_{\rm min})\approx n^0_f({\bf r}_{\rm max})+n^0_f({\bf r}_{\rm
min})$ in denominator. To directly see the image of the fermion
cloud, one may use a magnetic field to separate the fermion cloud
from boson cloud  before recording the fermion image in the
time-of-flight as splitting components in a spinor Bose atom
condensate \cite{sp}.

The Fermi surface of pure cold fermions has been observed in a
recent experiment \cite{kms} by the time-of-flight experiment.
When the mixture in the UMF state, as we have discussed in the
previous paragraph, $n_f({\bf r})=(\frac{m_f}t)|\tilde w_f({\bf
k})|^2n_{c}^1({\bf k}=\frac{m_f{\bf r}}t)$. Therefore, it is
expected that instead of the Fermi surface of the free Fermi
atoms, one may observe the 'electron' Fermi surface of the
'electron' gas. Namely, in the image of the time-of-flight, most
of Fermi atoms are inside of area with $|{\bf r}|<|{\bf
k}_F^1|t/m_f$ but not $|{\bf r}|<|{\bf k}_F|t/m_f$ ( $|{\bf k}_F|$
is Fermi surface of the free Fermi atoms.).

In conclusions, we showed that there may be electron-like and
photon-like excitations in  mixtures of Bose-Fermi atoms in
optical lattices in which the bosons are in the MI phase ($U_{BB}$
and $U_{BB}-U_{BF}\gg t_B$) and the fermions are nearly  half
filling. (To avoid the demixing, $4\pi t_F\sin (\pi n^f) U_{BB}>
U_{BF}^2$.) It was suggested that through the time-of-flight
experiment, the electron-like response function and the 'electron'
Fermi surface may be measured. An electron-like response function
also implies a gauge field effect. However, an experiment how to
directly observe a 'photon' was not designed yet.

This work was supported in part by Chinese National Natural
Science Foundation and the NSF of USA.


\begin{references}

\bibitem{exa} For examples, see, C. Orzel, A. K. Tuchman, M. L. Fenselau, M. Yasuda, and M. A. Kasevich
Science \textbf{291} 2386 (2001).M. Greiner, O. Mandel, T.
Esslinger, T. W. H\"{a}nsch, and I. Bloch, Nature (London)
\textbf{415}, 39 (2002). D. Jaksch, C. Bruder, J. I. Cirac, C. W.
Gardiner and P. Zoller, Phys. Rev. Lett. \textbf{81}, 3108 (1998).

\bibitem{rbk} G. Modugno et al, Phys. Rev. A {\bf 68}, 011601(R) (2003).  C. Schori et al, Phys. Rev. Lett {\bf 93}, 240402 (2004).  S.
Inouye et al, Phys. Rev. Lett. {\bf 93}, 183201 (2004). J. Goldwin
et al, Phys. Rev. A {\bf 70}, 021601(R) (2004).

\bibitem{nali} C. A. Stan et al,  Phys. Rev. Lett. {\bf 93} 143001
(2004).


\bibitem{aie} A. Albus, F. Illuminati, J. Eisert, Phys. Rev. A {\bf 68},
023606 (2003).

\bibitem{wen} For emergence of 'Photon' and 'electron' from
a non-relativistic model, see, e.g., P. A. Lee, N. Nagaosa, X. G.
Wen, cond-mat/0410445 and references therein.

\bibitem{comp} M. Lewenstein, L. Santos, M. A. Baranov, and H.
Fehrmann, Phys. Rev. Lett. {\bf 92}, 050401 (2004).



\bibitem{change} C. Schori, T. St\"oferle, H. Moritz, M. K\"ohl,
and T. Esslinger, Phys. Rev. Lett. {\bf 93}, 240402 (2004).



\bibitem{kms} M. K\"ohl, H. Moritz, T. St\"oferle, K. G\"unter, and
T. Esslinger, Phys. Rev. Lett. {\bf 94}, 080403 (2005).




\bibitem{d} D. B. Dickerscheid, D. van Oosten, P.
J. H.¡¡Densteneer, and H. T. C. Stoof, Phys. Rev. A {\bf 68},
043623(2003).

\bibitem{yu} Yue Yu and S. T. Chui, Phys. Rev. A {\bf 71},
033608(2005).

\bibitem{lu}  X. C. Lu, J. B. Li, and Y. Yu, cond-mat/0504503.


\bibitem{tj}  I. Affleck and J. B. Marston,
Phys. Rev. B {\bf 37}, 3774 (1988).


\bibitem{il} L. B. Ioffe and A. I. Larkin,
Phys. Rev. B {\bf 39}, 8988 (1989).


\bibitem{vi} F. Gerbier, A.
Widera, S. F\"olling, O. Mandel, T. Gericke, and I. Bloch, Phys.
Rev. Lett. {\bf 95}, 050404 (2005).


\bibitem{sp} H.J. Miesner, D.M. Stamper-Kurn, J. Stenger, S. Inouye,
A.P. Chikkatur and W. Ketterle, Phys. Rev. Lett. {\bf 82}, 2228
(1999).
\end{references}
\end{document}